# Relationship between blood pressure and flow rate in arteries using a modified Windkessel model

Nam Lyong Kang[*]

[*] Department of Nanomechatronics Engineering, Pusan National University, Miryang 50463, Republic of Korea

This study examined the flow rate in arteries using the modified Windkessel model, considering various models for blood pressure. An exact solution was derived using a Laplace transform method and the effects of blood pressure on the flow rate in an artery were examined. The effects of the flow resistance, arterial compliance, and inertia of blood on the flow rate were also investigated. The flow rate decreased with increasing inertia of the blood and flow resistance and decreasing arterial compliance. The height and position of the secondary peak were determined by a combination of the flow resistance, arterial compliance, and blood inertance. The results suggest that the risk of hypertension may increase with age because decreases in flow rate due to an increase in flow resistance and a decrease in arterial compliance were more substantial than the increase in flow rate caused by a decrease in blood inertance. The proposed method can provide information to examine how factors, such as aging, disease, and exercise, affect the flow rate in blood vessels.

# Introduction

The temporal behavior of blood flowing out of the heart is related to blood pressure and depends on the elasticity of the aorta and peripheral vascular resistance. The characteristics of this system can be formulated using an analogy with the formalism of the electric current in the Windkessel model. In this model, Poiseuille's law for the pressure drop in a tube corresponds to Ohm's law for the voltage drop in a resistor. Therefore, for an aorta and artery without compliance, the Poiseuille's formula determines the characteristics of blood flow. In this case, all the blood flowing out of the heart reaches the capillary tubes through the peripheral vascular system. On the other hand, some of the blood is stored around the artery walls if the aorta and artery are somewhat compliant. In a healthy system, the aorta and artery should be fully elastic to allow blood to flow in and out of the heart in the diastolic stage and systolic stage, respectively.

The original Windkessel model developed by Otto Frank [1] has two elements and represents the heart as a source of flow that pumps blood into the arterial system. The heart is lumped into two elements: the peripheral vascular resistance and arterial compliance. The individual resistances and compliances of all vessels are added in the Windkessel model. Therefore, resistance and compliance are called peripheral resistance and total compliance, respectively. On the other hand, the two-element Windkessel model produces unrealistic aortic pressure wave properties when the aortic flow is used as input [2]. To overcome this weakness of the two-element Windkessel model, the model was later modified to include the characteristic impedance and inertia of the blood [3-7]. In these modified models, the blood pressure, flow rate, peripheral resistance, arterial compliance, and blood inertance correspond to the electric voltage, electric current, electric resistance, capacitance, and self-inductance, respectively. The characteristic impedance links the Windkessel model to the wave travel aspects of the artery system. The proximal compliance, distal compliance, and peripheral resistance are connected in parallel.

On the other hand, the aging process decreases the arterial compliance and increases the peripheral resistance because with age, the stiffness of the artery wall increases, the viscosity increases, and the radius of the artery wall decreases [8]. In this case, the compliance and inertia of blood play important roles in the flow rate. The aging process also affects the systolic rising edge (SRE) and the peak position of the blood pressure because the shape characteristics of blood pressure are affected by the increasing arterial stiffness with age [8-11]. Zahedi et al. [8] showed that a steep rise and early peak of blood pressure occurred in younger subjects. The rise was quite gradual and the pulse peak appeared much later in older subjects, i.e., the peak time($T_p$) taken to reach the peak of the blood pressure after mitral valve closure is longer in older people than in younger people. This shows that the vascular parameters change with age, and the SRE could be considered a parameter in determining the age-dependent vascular state. The arterial stiffness is also affected by elevated blood pressure and higher sympathetic activities.

This study examined the behavior of circulatory blood flow inside the human body using an analogy with an electric circuit system. For this purpose, an exact solution for the modified Windkessel model was derived by applying the Laplace transform method, which is a powerful tool for obtaining a series solution for differential equations when a periodic driving force is applied. The flow rates calculated using the models for blood pressure introduced in this paper were fitted to actual in vivo results to confirm the validation of the method. The effects of the slope of the SRE and the peak time on the flow rate in an artery were examined using various models for blood pressure. The effects of the peripheral resistance, arterial compliance, and inertia of blood on the flow rate in the artery were then investigated. The risk of hypertension with age is also discussed.

**Methods**

Adapting the well-known Windkessel models [3-5], the main artery system (modified Windkessel model) shown in Fig. 1 was considered to examine the effects of blood pressure on the flow rate. The peripheral resistance, $R$, represents the resistance to blood flow in the arterial system, which is mainly in the resistance vessels, i.e., small arterials and arterioles. The resistance is the relationship between the pressure drop across a vessel and the blood flow through it and can be quantified by Poiseuille's law. The law relates the resistance to the length and radius of a vessel and the viscosity of blood under the assumption that the flow is steady and the vessel is rigid and uniform. The arterial compliance of the arterial wall, $C$, is the ratio of a volume change to the resulting change in pressure of a blood vessel that consists of large elastic arteries and small distal arteries. The compliance depends on the radius and wall thickness of the vessel and will decrease with age due to progressive changes in the elastin and collagen content of the arterial wall. The blood inertance, $L$, represents the inertia of blood mass in the artery.

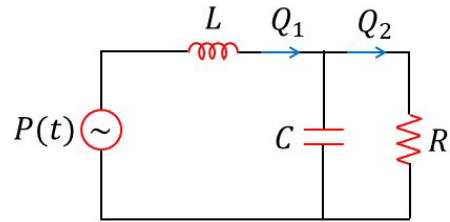

**Fig. 1** Modified Windkessel model for the human arterial system. $L$ is the blood inertance, $C$ is the arterial compliance, and $R$ is the peripheral resistance. $Q_1$ and $Q_2$ are the flow rates in the artery and peripheral blood vessels, respectively.

The blood flow increases dramatically at the beginning of the systolic state, and the human body can maintain steady blood flow. Therefore, it is reasonable to include an inertial element [3]. In previous models, the peripheral resistance is the main factor for the risk of hypertension, but arterial compliance is also important for old-age systolic hypertension [12, 13]. The blood inertance is predominant in large arteries, where the resistive effects do not play an important role. To examine the effects of the slope of the SRE and the peak time ($T_p$) on the flow rate in the artery, this paper considers the periodic blood pressures with the same period ($T = 0.85$ [s]) given in Table 1 and does not consider the dicrotic notch (DN) in the blood pressure (this will be done in a subsequent study). From the results reported by Zahedi et al. [8], models 1, 2, and 3-6 corresponds to younger subjects, middle-age subjects, and older subjects, respectively.

Models 1-3 (Fig. 2) have the same $L, R,$ and $C$ values as well as the same systolic and diastolic pressures, but their peak times and SREs are different. Models 3-6 (Fig. 3) have the same peak time but their systolic and diastolic pressures are different. Models 4-6 have the same $L, R,$ and $C$ values. The age-dependence of the blood flow rate was investigated using models 1-3, and the systolic pressure- and diastolic pressure-dependences of the blood flow rate were investigated using models 3-6.

**Table 1** Models for blood pressure in the artery considered in this paper.

($L$: mmHg·s$^2$/ml, $C$: ml/mmHg, $\gamma$: mmHg·s/ml)

| blood pressure : $P(t) = h_1 + h_2(t + h_3)\,e^{-gt} = P(t + T)$ [mmHg] | | | | | | | $L, R, C$ |
|---|---|---|---|---|---|---|---|
| Model | $h_1$ [mmHg] | $h_2$ [mmHg/s] | $h_3$ [s] | $g$ [1/s] | $T_p$ [s] | min. | max. | |
| 1 | 78 | 790 | 0.0022 | 7 | 0.014 | 80 | 120 | $L = 0.015$ |
| 2 | 72 | 615 | 0.0123 | 5 | 0.019 | | | $C = 0.15$ |
| 3 | 43 | 507 | 0.0735 | 3 | 0.026 | | | $R = 1.0$ |
| 4 | 62 | 380 | 0.0735 | 3 | 0.026 | 90 | 120 | $L = 0.0145$ |
| 5 | 33 | 635 | | | | 80 | 130 | $C = 0.145$ |
| 6 | 53 | 507 | | | | 90 | 130 | $R = 1.05$ |

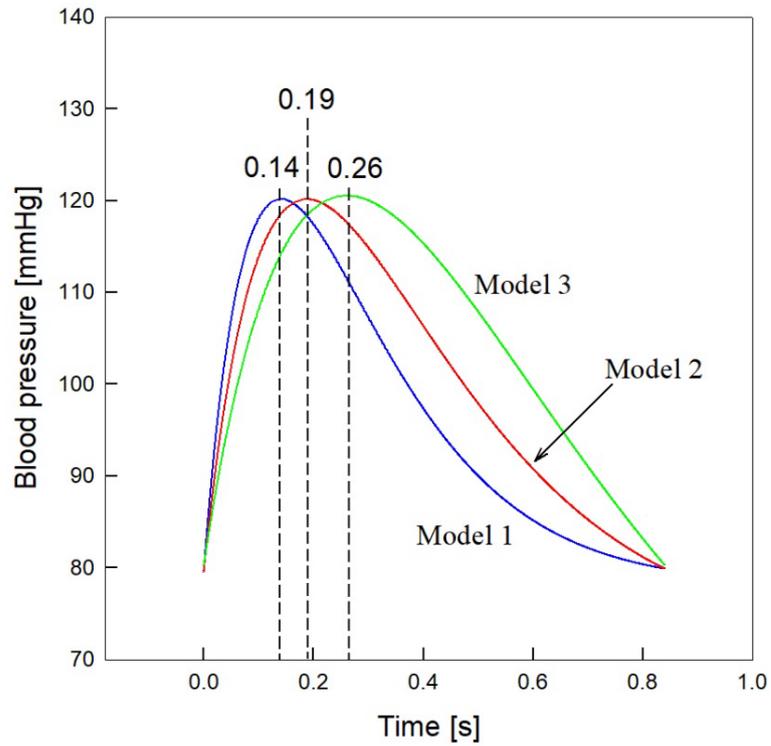

**Fig. 2** Models 1 – 3 for blood pressure with the same period, $T = 0.85$ [s]. The models are distinguished by their peak times, $0.14$ [s] for model 1, $0.19$ [s] for model 2, $0.26$ [s] for model 3.

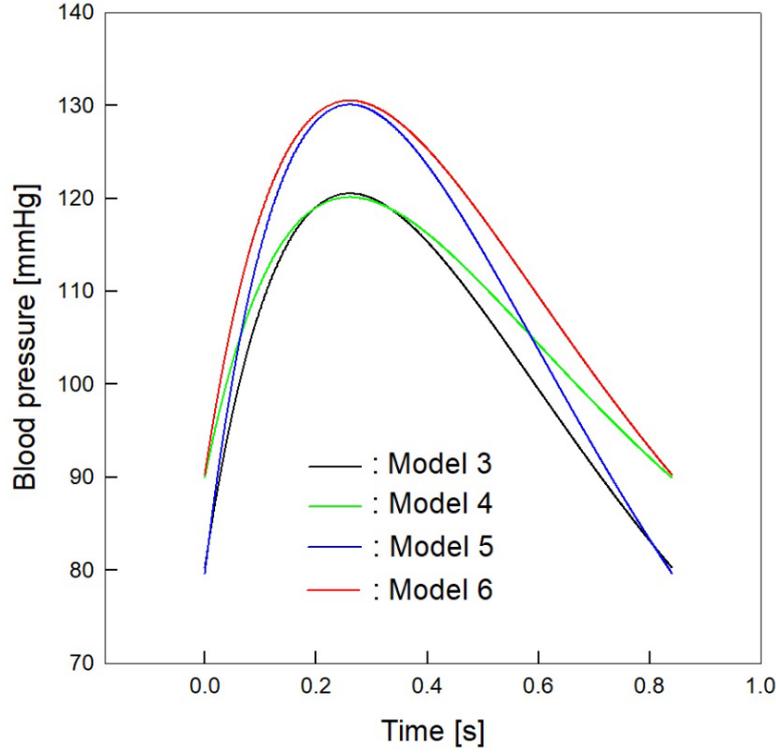

**Fig. 3** Models 3 - 6 for blood pressure with the same peak time and same period, $T_p = 0.26$ [s] and $T = 0.85$ [s]. The models are distinguished by their pressure ranges, 80 ~ 120 [mmHg] for model 3, 90 ~ 120 [mmHg] for model 4, 80 ~ 130 [mmHg] for model 5, and 90 ~ 130 [mmHg] for model 6.

The equation for blood flow can be derived by applying Kirchhoff's law to the artery system in Fig. 1, as follows:

$$\frac{dQ_1}{dt} = -aQ_2 + f(t), \qquad \frac{dQ_2}{dt} = bQ_1 - bQ_2 \tag{1}$$

where $a \equiv R/L, b \equiv 1/RC, f(t) = P(t)/L$. By applying the Laplace transform to Eq. (1), the Laplace transform of $Q_1$, $\tilde{Q}_1$, can be expressed as follows:

$$\tilde{Q}_1 = \frac{g_1(s+b)}{s(s^2+bs+ab)} + \frac{g_2 h_3(s+b)}{(s+\gamma)(s^2+bs+ab)(1-e^{-sT})}$$
$$- \frac{g_2(T+h_3)(s+b)e^{-(s+\gamma)T}}{(s+\gamma)(s^2+bs+ab)(1-e^{-sT})} + \frac{g_2(s+b)[1-e^{-(s+\gamma)T}]}{(s+\gamma)^2(s^2+bs+ab)(1-e^{-sT})} \tag{2}$$

where $g_1 \equiv h_1/L, g_2 \equiv h_2/L$, and $s$ is the Laplace variable. $Q_1(0)$ is neglected to consider the steady state flow rates. Expanding Eq. (2) into partial fractions results in the following:

$$\tilde{Q}_1 = g_1 \left[ \frac{A_1(s+b/2) + A_2 - A_1 b/2}{(s+b/2)^2 + \left(\sqrt{ab - b^2/4}\right)^2} - \frac{A_1}{s} \right]$$

$$+ g_2 \left[ \frac{B_1(s+b/2) + B_2 - B_1 b/2}{(s+b/2)^2 + \left(\sqrt{ab - b^2/4}\right)^2} - \frac{B_1}{s+\gamma} \right]$$

$$\times \left[ h_3 \sum_{n=0}^{\infty} e^{-nsT} - (T + h_3) e^{-\gamma T} \sum_{n=0}^{\infty} e^{-(n+1)sT} \right]$$

$$+ g_2 \left[ \frac{D_1}{s+\gamma} + \frac{D_2}{(s+\gamma)^2} - \frac{D_1(s+b/2) - D_3 + D_1 b/2}{(s+b/2)^2 + \left(\sqrt{ab - b^2/4}\right)^2} \right]$$

$$\times \left[ \sum_{n=0}^{\infty} e^{-nsT} - e^{-\gamma T} \sum_{n=0}^{\infty} e^{-(n+1)sT} \right] \qquad (3)$$

where the Taylor expansion, $(1-x)^{-1} = \sum_{n=0}^{\infty} x^n$, is used and

$$A_1 \equiv -\frac{1}{a}, \quad A_2 \equiv 1 - \frac{b}{a}, \quad B_1 \equiv \frac{\gamma - b}{\gamma(\gamma - b) + ab}, \quad B_2 \equiv \frac{(\gamma - b)b + ab}{\gamma(\gamma - b) + ab}$$

$$D_1 \equiv \frac{ab - \gamma^2 + b(2\gamma - b)}{H}, \quad D_2 \equiv \frac{(b - \gamma)(\gamma^2 - 2\gamma b) + b(ab + b\gamma - \gamma^2) - \gamma ab}{H}$$

$$D_3 \equiv \frac{(b - \gamma)b(b - a) + \gamma ab - b(ab + b\gamma - \gamma^2)}{H}$$

$$H \equiv (b - \gamma)(b\gamma^2 - 2ab\gamma) - (ab + b\gamma - \gamma^2)(\gamma^2 - ab) + \gamma ab(2\gamma - b) \qquad (4)$$

Therefore, applying the inverse Laplace transform to Eq. (3) results in

$$Q_1(t) = g_1 A_1 \left[ e^{-bt/2} \cos(\omega t) - 1 \right] + g_1 \frac{A_2 - A_1 b/2}{\omega} e^{-bt/2} \sin(\omega t)$$

$$+ g_2 (h_3 B_1 - D_1) \sum_{n=0}^{\infty} e^{-bt_n/2} \cos(\omega t_n) u(t_n) - g_2 (h_3 B_1 - D_1) \sum_{n=0}^{\infty} e^{-\gamma t_n} u(t_n)$$

$$+ g_2 \left( h_3 \frac{B_2 - B_1 b/2}{\omega} + \frac{D_3 + D_1 b/2}{\omega} \right) \sum_{n=0}^{\infty} e^{-bt_n/2} \sin(\omega t_n) u(t_n)$$

$$- g_2 e^{-\gamma t} [(T + h_3) B_1 - D_1] \sum_{n=0}^{\infty} e^{-bt_{n+1}/2} \cos(\omega t_{n+1}) u(t_{n+1})$$

$$- g_2 e^{-\gamma t} \left[ (T + h_3) \frac{B_2 - B_1 b/2}{\omega} + \frac{D_3 + D_1 b/2}{\omega} \right] \sum_{n=0}^{\infty} e^{-\frac{bt_{n+1}}{2}} \sin(\omega t_{n+1}) u(t_{n+1})$$

$$+g_2 e^{-\gamma t}[(T+h_3)B_1 - D_1] \sum_{n=0}^{\infty} e^{-\gamma t_{n+1}} u(t_{n+1})$$

$$+g_2 D_2 \sum_{n=0}^{\infty} t_n e^{-\gamma t_n} u(t_n) - g_2 D_2 e^{-\gamma T} \sum_{n=0}^{\infty} t_{n+1} e^{-\gamma t_{n+1}} u(t_{n+1}) \qquad (5)$$

where $u[x]$ is the step function, $\omega \equiv \sqrt{ab - b^2/4}$, and $t_n \equiv t - nT$.

## Results

Fig. 4 validates the present method. The colored lines indicate the flow rates in the artery calculated using the present method for models 1-3 at $T = 0.85[\text{s}]$ and are compared with the experimental results (black line) reported by Alastruey et al. [14] in the right iliac-femoral. The figure shows that the $Q_1$ values oscillate in the order of primary peaks ($P_1$) - dicrotic notches (DN) - secondary peak ($P_2$) and three models explain the characteristic features of the experimental results. The interval of $P_1$ and DN is shorter than the experiment result, and there is no plateau (e.g., at point A). Therefore, $P_1$ arises later, and DN arises earlier than the experiment. This discrepancy is expected to be improved if a realistic blood pressure, including the dicrotic notch in the blood pressure, is considered. The present results are the flow rates in the whole artery system, while the experiment result is the flow rate only at the right iliac-femoral. Therefore, the present results were rescaled by multiplying $1/26$ for comparison with the experimental result.

The DN and $P_2$ in Fig. 4 were attributed to the arterial compliances ($C$), i.e., the artery takes in blood and then resupplies it to the flow. $P_2$ in $Q_1$ arises because the flow stored in the artery is restored, while $P_2$ in $Q_2$ (flow rate in peripheral blood vessel) arises because the blood stored in the artery is resupplied to the peripheral blood vessels. Therefore, the primary and secondary peaks of $Q_1$ arise before that of $Q_2$, and the range of $Q_1$ is larger than that of $Q_2$ because of the buffer action of $C$. The figure shows that model 1 oscillates more deeply than model 3 because the height of $P_1$ and the depth of DN in model 1 are larger than those in model 3. This means that the oscillation becomes blunted with age because the peak time ($T_\text{p}$) of model 3 is longer than that of model 1, and the peak time increases with age [8]. The figure also shows that the risk of hypertension increases with age because the peak of $Q_1$ decreases with age. Therefore, the systolic blood pressure increases to supply more blood to the vessel.

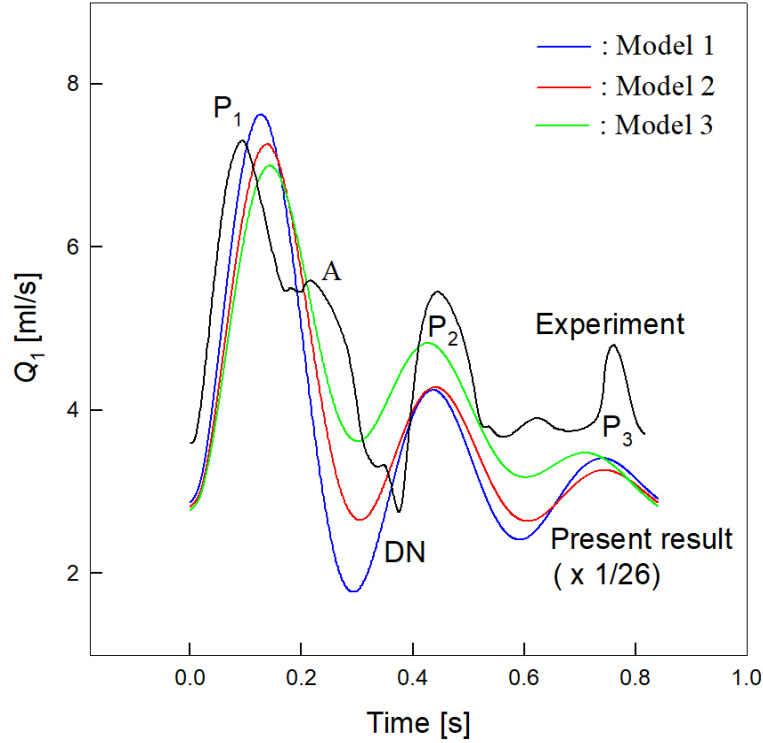

**Fig. 4** Flow rates in an artery. The colored lines indicate the flow rates in an artery calculated using the present method for models 1-3 at $T = 0.85$ [s]. The black line indicates the experimental result reported by Alastruey et al. [14] for the flow rate in the right iliac-femoral.

Figure 5 shows that the depth of DN and the height of $P_1$ in $Q_1$ for the model 2 at $T = 0.85$ [s] decrease as $C$ decreases, while $P_2$ is nearly independent of $C$. This occurs because less blood is stored in the artery as $C$ decreases. In Eq. (4), $\gamma - b = \gamma - 1/RC$ decreases and $\omega$ increases as $C$ decreases. Therefore, the height of the primary peak decreases and $P_2$ moves into the left primary peak as $C$ decreases because the terms $\cos[\omega(\cdots)]$ and $\sin[\omega(\cdots)]$ in Eq. (5) oscillate more quickly as $C$ decreases. This means that people with a small $C$ should refrain from intense exercise because sufficient blood cannot be supplied to the blood vessels, and they have a risk of hypertension.

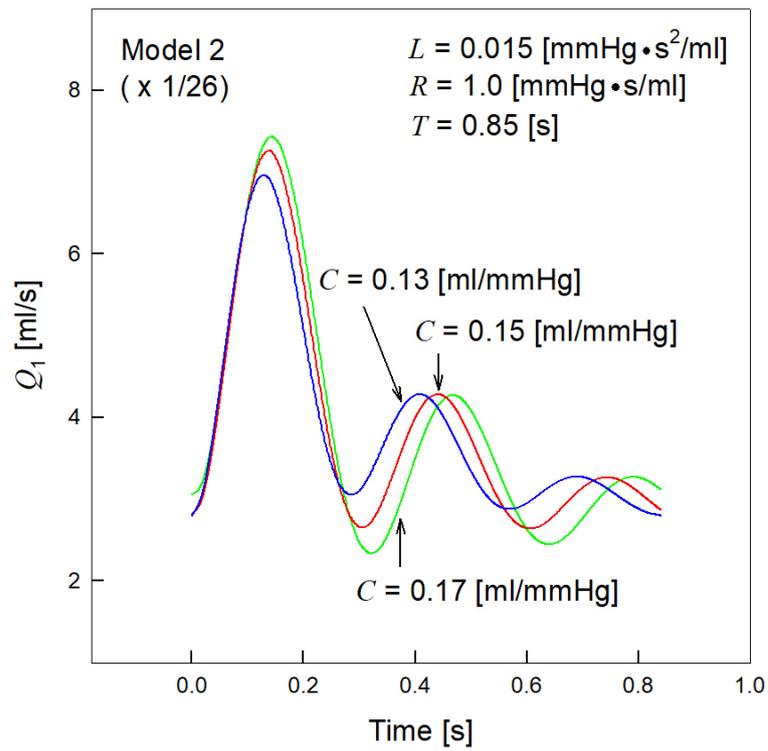

**Fig. 5** Flow rates in an artery of model 2 for various arterial compliances.

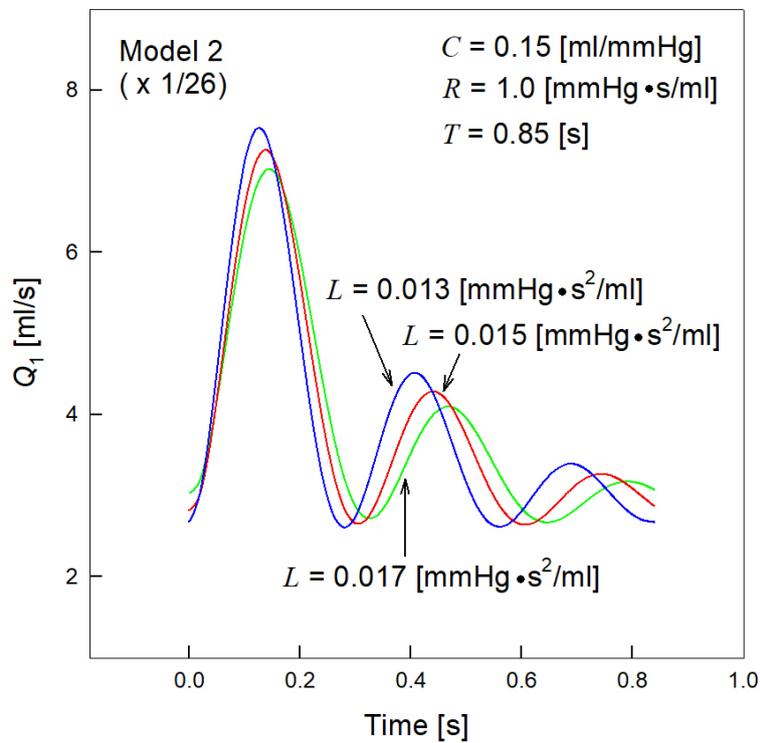

**Fig. 6** Flow rates in an artery of model 2 for various blood inertances.

Figure 6 shows that the height of $P_1$ in $Q_1$ for model 2 at $T = 0.85$ [s] decreases with increasing $L$, and the fluctuations increase. This is because the blood inertance tends to maintain the flow rate; thus, less blood is supplied to the artery as the blood inertance increases. Therefore, people with a large $L$ have a risk of hypertension because their lung should increase the systolic pressure to supply sufficient blood to the vessels. The height of $P_2$ decreases with increasing $L$ because the drop in blood pressure increases with increasing $L$ according to Faraday's law. The reason why $P_2$ moves into the left primary peak as $L$ decreases is the same as that in Fig. 5. Figure 7 shows that the flow rates for model 2 decrease with increasing peripheral resistance. Therefore, an increase in peripheral resistance may cause hypertension because the lung should increase the systolic pressure to supply sufficient blood to the vessels.

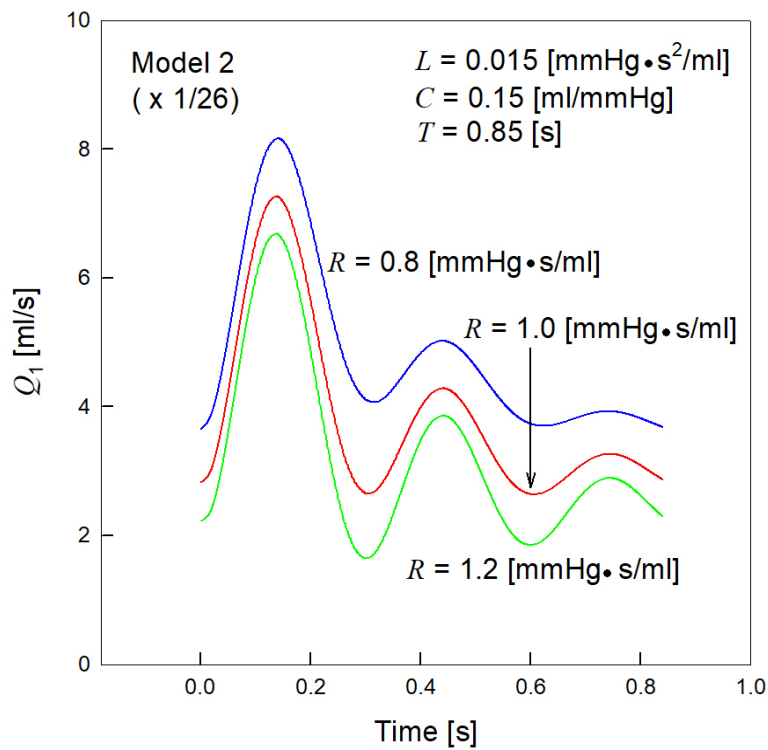

**Fig. 7** Flow rates in an artery of model 2 for various peripheral resistances.

Figures 5 and 6 show that the flow rate decreases with decreasing compliance ($C$), whereas it decreases with increasing blood inertance ($L$). This means that the flow rate is determined by $C$ and $L$ in a complicated manner. To investigate this further, the flow rates for model 3 were recalculated with different values of $C$ and $L$ (Fig. 8). The figure shows that the peak times (0.142 [s]) of models 4-6 were closer to the experimental data than that (0.149 [s]) of model 3. The primary peak of the flow rate was highest for model 5, whereas the third peak of the flow rate was highest for model 4. This means that the early flow rate is affected most by the systolic pressure, whereas the diastolic pressure determines the late flow rate. Therefore, model 6 fits the real data better than model 3. $L$ and $C$ of models 4-6 are smaller than those of model 3, and $R$ of models 4-6 is larger than that of model 3 (Table 1).

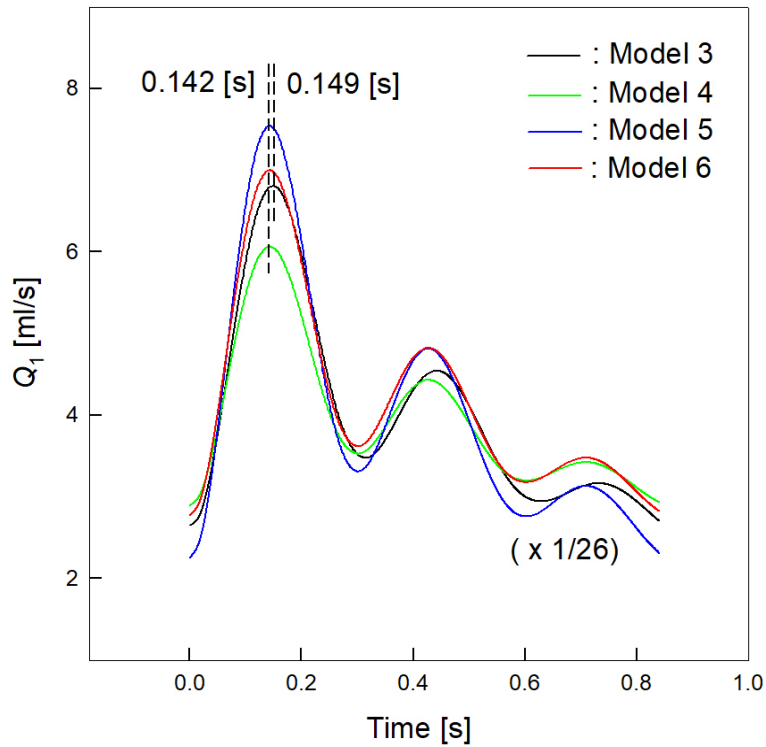

**Fig. 8**  Flow rates in an artery for different values of $C$, $L$, $R$, and blood pressure

The compliance and inertance decrease with age, but the resistance increases [10]. Figure 8 shows that the decrease in flow rate with decreasing $C$ and increasing $R$ is larger than the increase in flow rate with increasing $L$. Therefore, the risk of hypertension increases with age because the increases in systolic and diastolic pressures with decreasing $C$ and increasing $R$ to supply sufficient blood to the arteries with age must be larger than the decrease in these values with decreasing $L$. Based on these results, the decrease in flow rate among older people could be attributed to the age-related systemic increase in resistance and decrease in compliance.

## Discussion

An exact solution for the modified Windkessel model was derived using the Laplace transform method and considering various types of blood pressure. The age - and cardiovascular - parameter dependences of hypertension were examined using three kinds of blood pressures (models 1-3) distinguished by the peak time in the artery. The effects of the arterial compliance ($C$), blood inertance ($L$), and peripheral resistance ($R$) on the flow rate in the artery were also investigated. The results explained that the artery takes up blood in the systolic stage and resupplies it to the flow in the diastolic stage. The height of the primary peak ($P_1$) and the depth of the dicrotic notch (DN) in the flow rate decrease with increasing peak time. The height of $P_1$ and the depth of DN decrease with increasing $L$ and $R$ but decrease with decreasing $C$.

In general, with age, the stiffness of the artery wall increases, the viscosity increases, and the radius of the artery wall decreases [8]. The arterial compliance decreases with increasing stiffness of the artery. The peripheral resistance increases with increasing viscosity, and the blood inertance decreases with decreasing radius of the artery wall. As a result, with age, the arterial compliance and blood inertance decrease, and the peripheral resistance increases. This paper explains why the risk of hypertension increases with age because, despite the decrease in blood inertance with age, the decrease in flow rate with increasing peripheral resistance and decreasing compliance is larger than the increase in flow rate with decreasing blood inertance. Therefore, the blood pressure increases with age to supply more blood to the artery. During exercise, the heart beats more frequently, and the pressure of the heart increases to supply more blood to the blood vessels in a short time because exercise shortens the period between heartbeats. This study showed that a person with a small $C$, large $L$, and large $R$ should limit intense exercise because they cannot supply sufficient blood to the blood vessels.

## Conclusion

In conclusion, the present method can be applied to determining the age-dependent vascular state, and some useful information about the blood circulation in the cardiovascular system can be obtained from this method. The method can also be used to indicate the cardiovascular risk because the resistance, compliance, and inertance vary with aging, hypertension, diabetes, and other factors. The present method can be improved by considering a more accurate form for blood pressure, including the DN in blood pressure, and can be applied to the four-element Windkessel model, including the wave transmission phenomena [3, 4]. All these studies will be performed in the near future.